\documentclass[npg]{copernicus}
\usepackage{color}
\usepackage{graphicx}

\begin{document}

\title{Achieving Fast Reconnection in Resistive MHD Models via Turbulent Means}

\author[1]{Giovanni Lapenta}
\author[2]{A. Lazarian}

\affil[1]{Centrum voor Plasma-Astrofysica, Departement Wiskunde, Katholieke Universiteit Leuven, Celestijnenlaan 200B, 3001 Leuven, Belgium (EU)}
\affil[2]{Department of Astronomy, University of Wisconsin-Madison, 475 Charter St., USA}


\runningtitle{ Fast Turbulent Reconnection}

\runningauthor{Lapenta and Lazarian}

\correspondence{Giovanni Lapenta\\ (giovanni.lapenta@wis.kuleuven.be)}

\received{}
\pubdiscuss{} 
\revised{}
\accepted{}
\published{}


\firstpage{1}

\maketitle

\abstract{Astrophysical fluids are generally turbulent and this preexisting turbulence
must be taken into account for the models of magnetic reconnection which are
attepmted to be applied to astrophysical, solar or heliospheric environments. 
In addition, reconnection itself induces turbulence which provides an important
 feedback on the reconnection process. In this paper we discuss both theoretical
model and numerical evidence that magnetic reconnection gets fast in the
approximation of resistive MHD. We consider the relation between the Lazarian \& Vishniac turbulent reconnection theory and Lapenta's numerical experiments 
testifying of the spontaneous onset of turbulent reconnection in systems which are 
initially laminar.

 }

\introduction

~~Astrophysical plasmas are known to be magnetized and turbulent. Magnetization of these fluids most frequently arises from the dynamo action to which turbulence is an essential component (see Schekochihin et al. 2007). The drivers of turbulence, e.g. supernovae explosions in the interstellar medium, inject energy at large scales and then the energy cascades down to small scales through a hierarchy of eddies spanning up over the entire inertial range. The famous Kolmogorov picture (Kolmogorov 1941) corresponds to hydrodynamic turbulence, but, as we discuss further, a qualitatively similar turbulence also develops in magnetized fluids/plasmas.
The definitive confirmation of turbulence presence comes from observations, e.g. observations of electron density fluctuations in the interstellar medium, which produce a so-called Big Power Law in the Sky (Armstrong et al. 1994, Chepurnov \& Lazarian 2010), with the spectral index coinciding with the Kolmogorov one. A more direct piece of evidence comes from the observations of spectral lines. Apart from showing non-thermal Doppler broadening, they also reveal spectra of supersonic turbulent velocity fluctuations when analyzed with techniques like Velocity Channel Analysis (VCA) of Velocity Coordinate Spectrum (VCS) developed (see Lazarian \& Pogosyan 2000, 2004, 2006, 2008) and applied to the observational data (see Padoan et al. 2004, 2009, Chepurnov et al. 2010) rather recently.

Reconnection is a process of changing magnetic flux topology and this process has been a challenge to understand in highly conductive astrophysical fluids. Naturally,
it is important to understand the process of magnetic reconnection in reallistically 
turbulent state of the fluid. Observationally, it is also known that a high level of 
turbulence is present as outcome of solar flares, which are generally believed to 
be driven by magnetic reconnection. Therefore it is important to understand the
mutual feedback of turbulence and reconnection and provide the connection
between the existing theory, observations and numerical experiments. This is the goal of the present paper. 

We consider both the model of turbulent reconnection suggested in Lazarian \& Vishniac (1999, henceforth LV99) and numerical evidence of fast reconnection in a few numerical papers by Lapenta and coauthors. We search for the relation of the theory and observations.

The very idea that turbulence can influence magnetic reconnection is not new.
However, LV99 model is radically different from its predecessors which also appealed  to the effects of turbulence. For instance, unlike Speiser (1970) and 
Jacobson (1984) the model does not appeal to changes of the microscopic properties of the plasma. 
The nearest progenitor to LV99 was the work of  Matthaeus \& Lamkin (1985, 1986, henceforth ML),  who suggested that magnetic reconnection 
may be fast due to a number of  turbulence effects, e.g. multiple X points and turbulent EMF. However,  ML did not realize the key role of played by magnetic field-line wandering\footnote{Instead a discussion other effects, like heating, was presented. Those do not play a role in LV99.}, and did not obtain a quantitative prediction for the reconnection rate, as did LV99. From the numerical
simulations in ML obtaining the actual reconnection rate is not straightforward. They
do not discuss the onset of reconnection from the laminar case, either. On the contrary, we shall demonstrate the results of simulations where both of these effects are studied and quantified. 

In what follows, we discuss magnetic reconnection in turbulent fluid in \S 2, provide
the numerical confirmations of the predictions of the LV99 model in \S3, consider
the spontaneous onset of reconnection in MHD simulations in \S 4 and the role
of flow pattern in \S 5. We discuss the implications of the expected flares of 
reconnection in \S 6 and provide our summary in \S 7. 

\section{Reconnection can be Fast in Turbulent Plasmas}

Astrophysical plasmas are often highly ionized and highly magnetized (Parker
1970).  The evolution of the magnetic field in a highly conducting fluid can be
described by a simple version of the induction equation
\begin{equation}
\frac{\partial \vec{B}}{\partial t} = \nabla \times \left( \vec{v} \times \vec{B} - \eta \nabla \times \vec{B} \right) ,
\end{equation}
where $\vec{B}$ is the magnetic field, $\vec{v}$ is the velocity field, and
$\eta$ is the resistivity coefficient.  Under most circumstances this is
adequate for discussing the evolution of magnetic field in an astrophysical
plasma.  When the dissipative term on the right hand side is small, as is
implied by simple dimensional estimates, the magnetic flux through any fluid
element is constant in time and the field topology is an invariant of motion.
On the other hand, reconnection is observed in the solar corona and chromosphere
(Innes et al. 1997, Yokoyama \& Shibata 1995, Masuda et al. 1994, Ciaravella \&
Raymond 2008), its presence is required to explain dynamo action in stars and
galactic disks (Parker 1970, 1993), and the violent relaxation of magnetic
fields following a change in topology is a prime candidate for the acceleration
of high energy particles (de Gouveia Dal Pino \& Lazarian 2003, henceforth GL03,
2005, Lazarian 2005, Drake et al. 2006, Lazarian \& Opher 2009, Drake et al.
2010) in the universe.   Quantitative general estimates for the speed of
reconnection start with two adjacent volumes with different large scale magnetic
field directions (Sweet 1958, Parker 1957).

\begin{figure}[!t]
 \begin{center}
\includegraphics[width=1.0 \columnwidth]{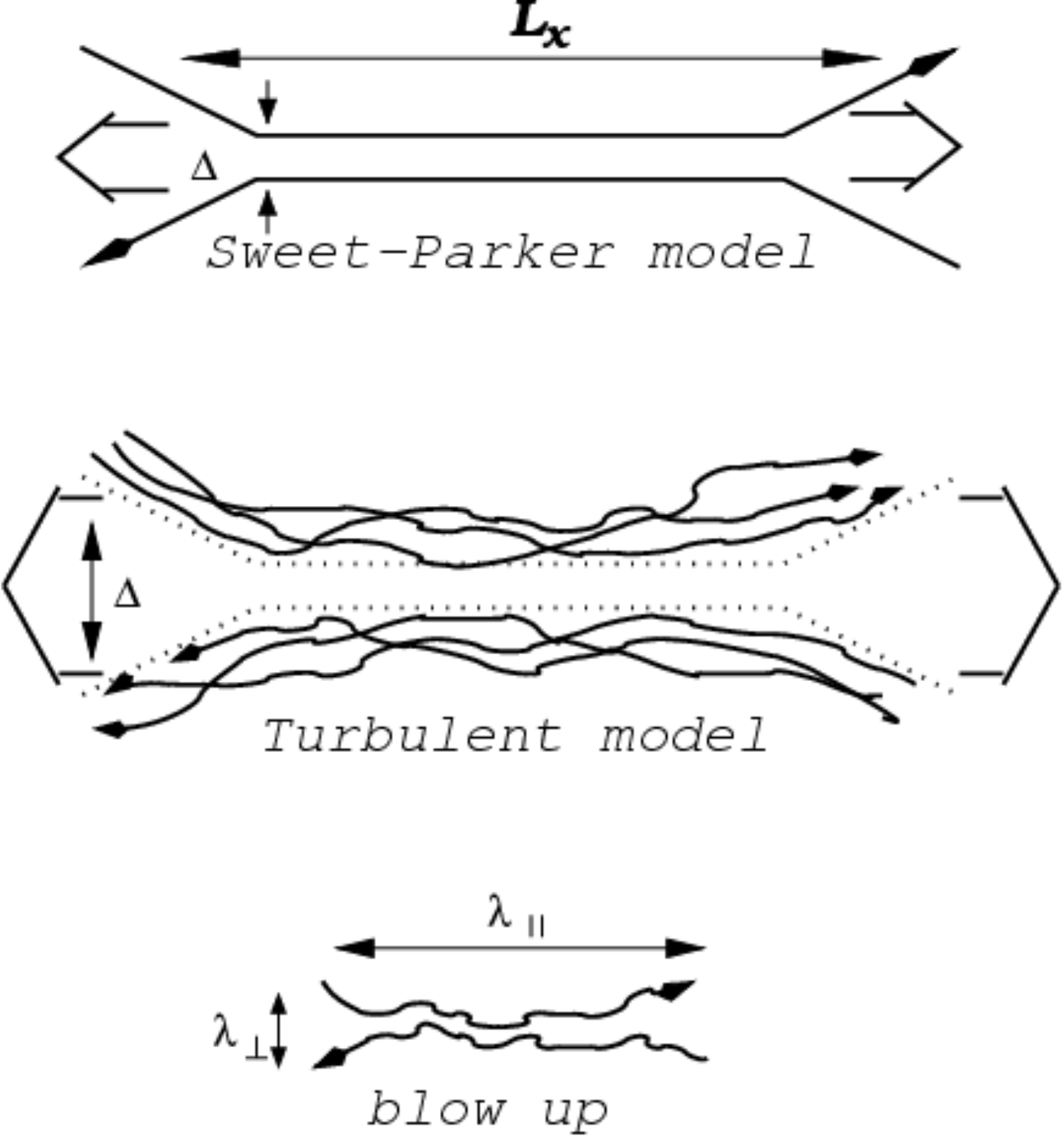}
\caption{{\it Upper plot}:
Sweet-Parker model of reconnection. The outflow is limited by a thin slot
$\Delta$, which is determined by Ohmic diffusivity. The other scale is an
astrophysical scale $L\gg \Delta$.
{\it Middle plot}: Reconnection of weakly stochastic magnetic field according to
LV99. The model that accounts for the stochasticity of magnetic field lines. The
outflow is limited by the diffusion of magnetic field lines, which depends on
field line stochasticity.
{\it Low plot}: An individual small scale reconnection region. The reconnection
over small patches of magnetic field determines the local reconnection rate. The
global reconnection rate is substantially larger as many independent patches
come together. From Lazarian et al. 2004.}
\label{fig_rec}
 \end{center}
\end{figure}

LV99 we introduced a model that included the effects of magnetic field line
wandering (see Figure 1). The model relies on the nature of three-dimensional
magnetic field wandering in turbulence. This nature is different in three and
two dimensions, which provides the major difference between the LV99 model and
the earlier attempts to solve the problem of magnetic reconnection appealing to
turbulence (Matthaeus \& Lamkin 1985). The effects of compressibility and
heating which were thought to be important in the earlier studies (Matthaeus \&
Lamkin 1985, 1986) do not play the role for the LV99 model either. The model is
applicable to any weakly perturbed magnetized fluid, irrespectively, of the
degree of plasma being collisional or collisionless (cf. Shay et al. 1998).

Two effects are the most important for understanding of the nature of
reconnection in LV99. First of all, in three dimensions bundles of magnetic
field lines can enter the reconnection region and reconnection there
independently (see Figure~1), which is in contrast to two dimensional picture
where in Sweet-Parker reconnection the process is artificially constrained.
Then, the nature of magnetic field stochasticity and therefore magnetic field
wandering (which determines the outflow thickness, as illustrated in Figure~1)
is very different in 2D and the real 3D world (LV99). In other words, removing
artificial constraints on the dimensionality of the reconnection region and the
magnetic field being absolutely straight, LV99 explores the real-world
astrophysical reconnection.

Our calculations in LV99 showed that the resulting reconnection rate is limited
only by the width of the outflow region.  This proposal, called ``stochastic
reconnection'', leads to reconnection speeds close to the turbulent velocity in
the fluid.  More precisely, assuming isotropically driven turbulence
characterized by an injection scale, $l$, smaller than the current sheet length,
we find
\begin{equation}
V_{rec}\approx \frac{u_l^2}{V_A}\left(l/L\right)^{1/2}\approx u_{turb}\left(l/L\right)^{1/2},
\label{recon1}
\end{equation}
where $u_l$  is the velocity at the driving scale and $u_{turb}$ is the velocity
of the largest eddies of the strong turbulent cascade. Note, that here "strong"
means only that the eddies decay through nonlinear interactions in an eddy turn
over time (see more discussion of the  LV99).  All the motions are weak in the
sense that the magnetic field lines are only weakly perturbed.

It is useful to rewrite this in terms of the power injection rate $P$. As the
perturbations on the injection scale of turbulence are assumed to have
velocities $u_l<V_A$, the turbulence is weak at large scales. Therefore, the
relation between the power and the injection velocities are different from the
usual Kolmogorov estimate, namely, in the case of the weak turbulence $P\sim
u_l^4/(lV_A)$ (LV99). Thus we get,
\begin{equation}
V_{rec}\approx \left(\frac{P}{LV_A}\right)^{1/2} l,
\label{recon2}
\end{equation}
where $l$ is the length of the turbulent eddies parallel to the large scale
magnetic field lines as well as the injection scale.

The reconnection velocity given by equation (\ref{recon2}) is obtained in MHD limit and therefore it does not depend on resistivity or plasma effects. Therefore LV99 
model predicts that for sufficiently high level of
turbulence collisionless and collisional fluids should reconnect at
the same rate.

\section{Fast reconnection in Turbulent MHD Simulations}
%
\begin{figure*}
\center
\includegraphics[width=0.3\textwidth]{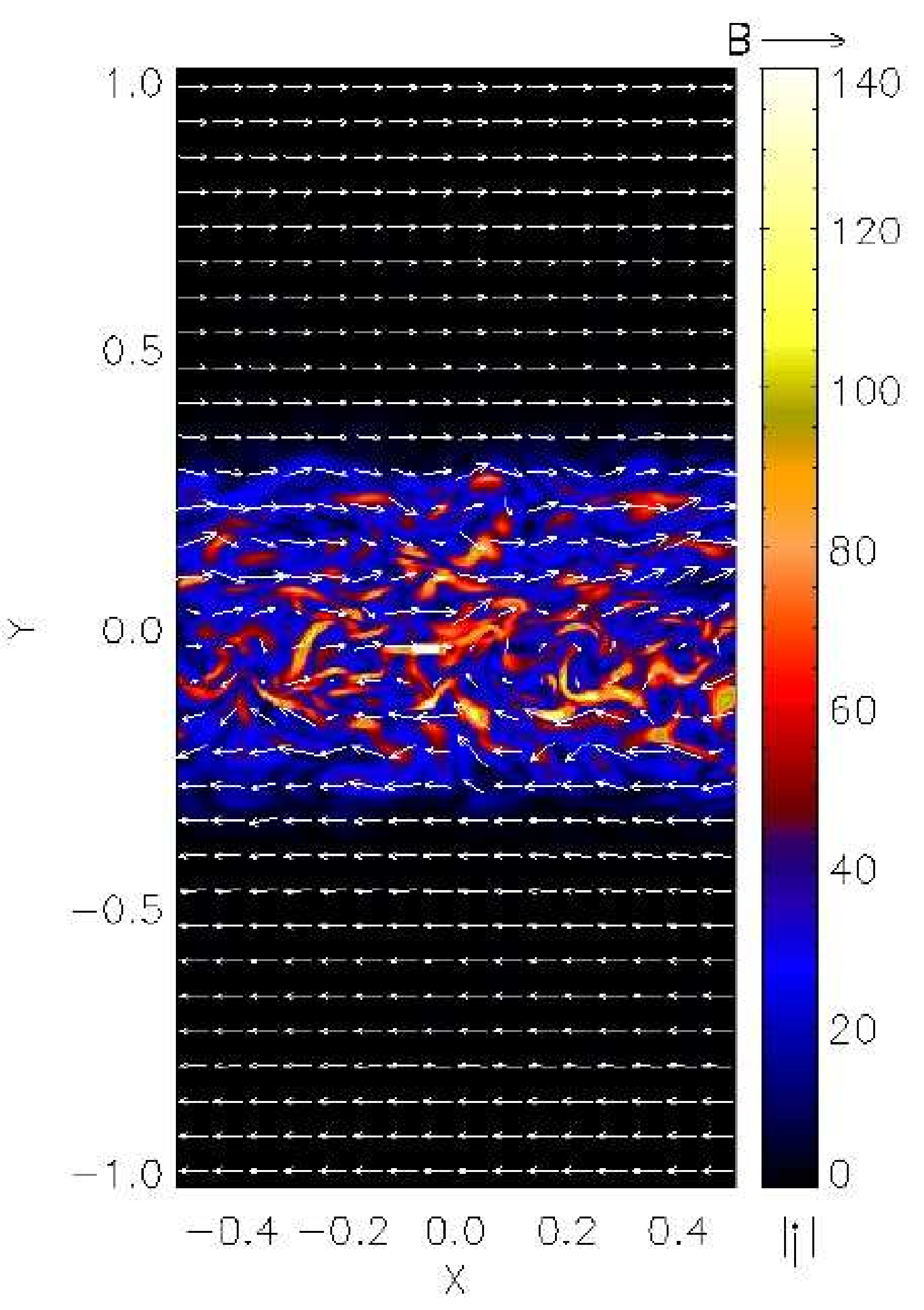}
\includegraphics[width=0.3\textwidth]{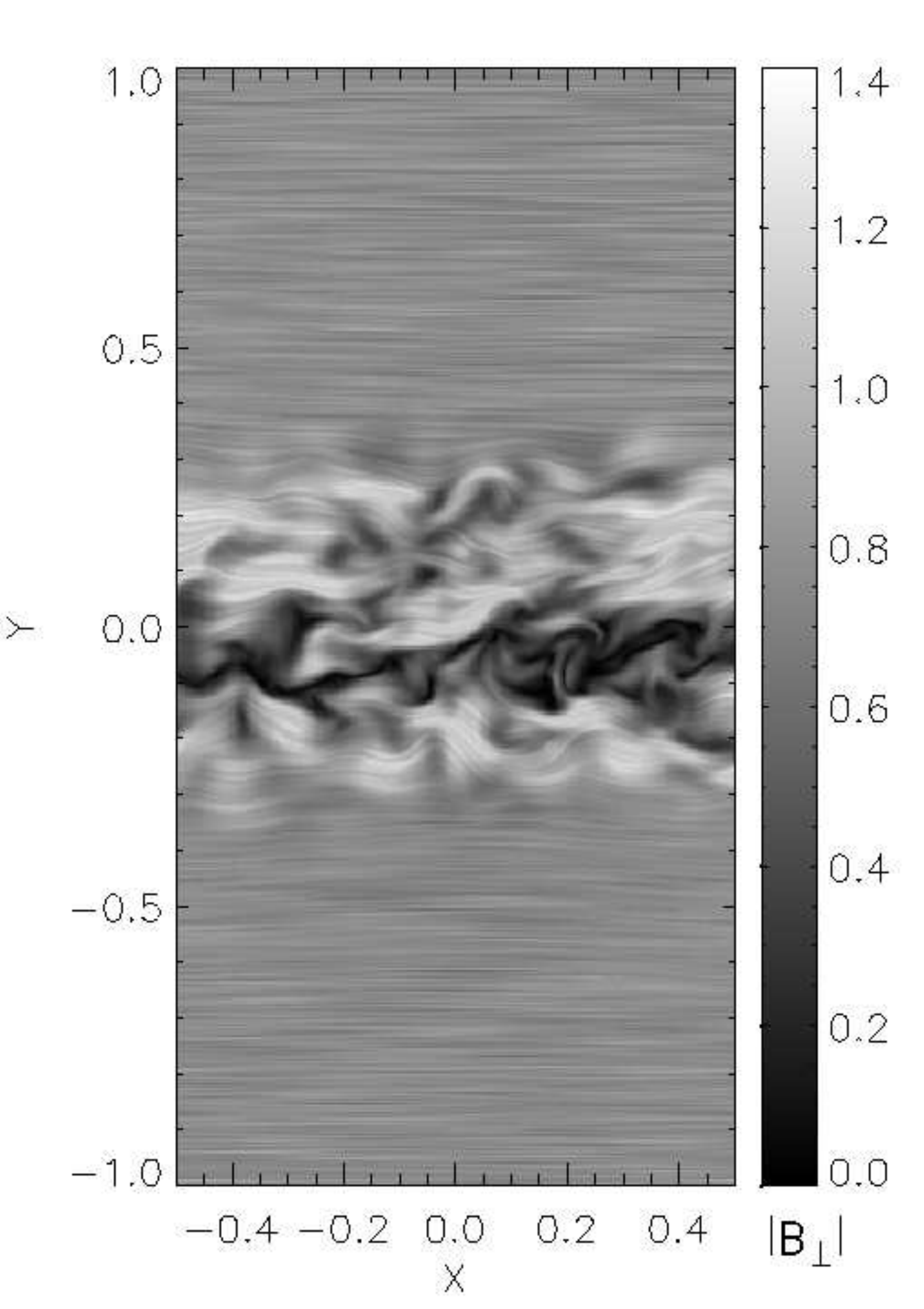}
\caption{{\it Left panel}: Current intensity and magnetic field configuration
during stochastic reconnection.  We show a slice through the middle of the
computational box in the xy plane after twelve dynamical times for a typical
run.  The shared component of the field is perpendicular to the page.  The
intensity and direction of the magnetic field is represented by the length and
direction of the arrows.  The color bar gives the intensity of the current.  The
reversal in $B_x$  is confined to the vicinity of y=0 but the current sheet is
strongly disordered with features that extend far from the zone of reversal.
{\it Right panel}: Representation of the magnetic field in the reconnection zone
with textures. \label{fig:top_turb}}
\end{figure*}

Here we describe the results of a series of three dimensional numerical
simulations aimed at adding turbulence to the simplest reconnection scenario and
testing equation (\ref{recon2}).  We take two regions with strongly differing
magnetic fields lying next to one another.  The simulations are periodic in the
direction of the shared field (the z axis) and are open in the reversed
direction (the x axis).  The external gas pressure is uniform and the magnetic
fields at the top and bottom of the box are taken to be the specified external
fields plus small perturbations to allow for outgoing waves.  The grid size in
the simulations varied from 256x512x256 to 512x1028x512 so that the top and
bottom of the box are far away from the current sheet and the region of driven
turbulence around it.   At the sides of the box where outflow is expected the
derivatives of the dynamical variables are set to zero.  A complete description
of the numerical methodology can be found in Kowal et al. (2009). All our
simulations are allowed to evolve for seven Alfven crossing times without
turbulent forcing.  During this time they develop the expected Sweet-Parker
current sheet configuration with slow reconnection.  Subsequently we turn on
isotropic turbulent forcing inside a volume centered in the midplane (in the xz
plane) of the simulation box and extending outwards by a quarter of the box
size.  The turbulence reaches its full amplitude around eight crossing times and
is stationary thereafter.

In Figure 2 we see the current density on an xy slice of the computational box
once the turbulence is well developed.  As expected, we see that the narrow
stationary current sheet characteristic of Sweet-Parker reconnection is replaced
by a chaotic structure, with numerous narrow peaks in the current density.
Clearly the presence of turbulence has a dramatic impact on the structure of the
reconnection zone.  In addition, we see numerous faint features indicating weak
reconnection between adjacent turbulent eddies.

\begin{figure}
\center
\includegraphics[width=0.9\columnwidth]{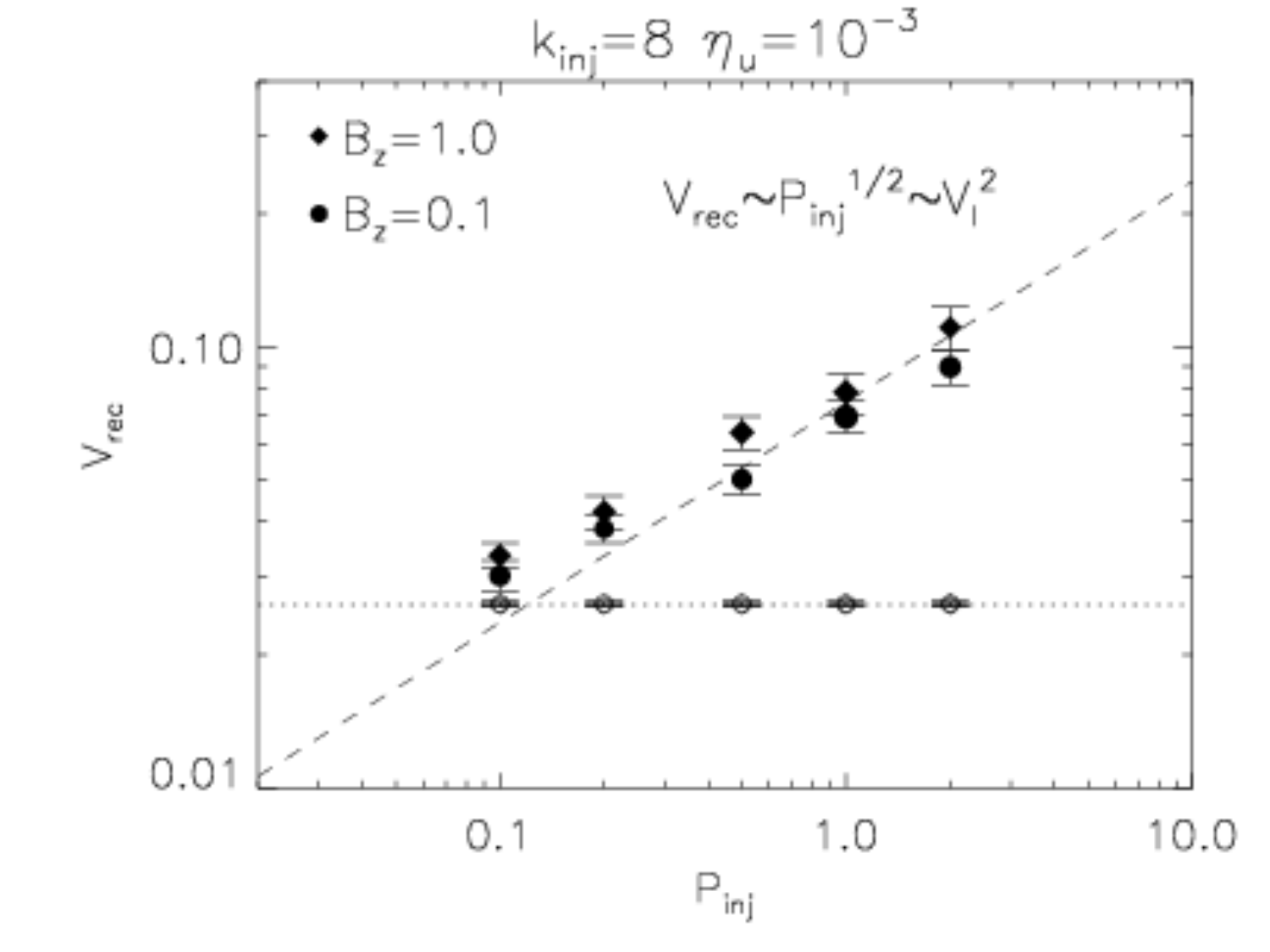}
\caption{Reconnection speed versus input power for the driven turbulence.  We
show the reconnection speed, defined by equation (4) plotted against the input
power for an injection wavenumber equal to 8 (i.e. a wavelength equal to one
eighth of the box size) and a resistivity $\nu_u$.  The dashed line is a fit to
the predicted dependence of  $P^{1/2}$ (see eq. (3)).  The horizontal line shows
the laminar reconnection rates for each of the simulations before the turbulent
forcing started.  Here the   uncertainty in the time averages are indicated by
the size of the symbols and the variances are shown by the error bars.
\label{pow_dep}}
\end{figure}

The speed of reconnection in three dimensions can be hard to define without
explicit evaluation of the magnetic field topology.  However, in this simple
case we can define it as the rate at which the $x$ component of the magnetic
field disappears.  More precisely, we consider a yz slice of the simulation,
passing through the center of the box.  The rate of change of the area integral
of  |$B_x$| is its flux across the boundaries of the box minus the rate at which
flux is annihilated through reconnection (see more discussion in Kowal et al.
2009)
\begin{equation}
\partial_t\left(\int|B_x|dzdy\right)=\oint sign(B_x)vec{E}d\vec{l}-2V_{rec}B_{x,ext}L_z
\label{measure}
\end{equation}
where electric field is $\vec{E}=\vec{v}\times \vec{B} -\eta \vec{j}$,
$B_{x,ext}$ is the absolute value of $B_x$  far from the current sheet and $L_z$
is the width of the box in the $z$ direction.  This follows from the induction
equation under the assumption that the turbulence is weak to lead to local field
reversals and that the stresses at the boundaries are weak to produce
significant field bending there.  In other words, fields in the $x$ direction
are advected through the top and bottom of the box, and disappear only through
reconnection.  Since we have assumed periodic boundary conditions in the $z$
direction the boundary integral on the right hand side is only taken over the
top and bottom of the box.  By design this definition includes contributions to
the reconnection speed from contracting loops, where Ohmic reconnection has
occurred elsewhere in the box and $|B_x|$ decreases as the end of a reconnected
loop is pulled through the plane of integration.  It is worth noting that this
estimate is roughly consistent with simply measuring the average influx of
magnetic field lines through the top and bottom of the computational box and
equating the mean inflow velocity with the reconnection speed. Following
equation (\ref{measure}) we can evaluate the reconnection speed for varying
strengths and scales of turbulence and varying resistivity.

In Figure~\ref{pow_dep} we see the results for varying amounts of input power,
for fixed resistivity and injection scale as well as for the case of no
turbulence at all.  The line drawn through the simulation points is for the
predicted scaling with the square root of the input power. The agreement between
equation (\ref{recon2}) and Figure~\ref{pow_dep} is encouraging but does not
address the most important aspect of stochastic reconnection, i.e. its
insensitivity to $\eta$.

%
\begin{figure}
\center
\includegraphics[width=0.9\columnwidth]{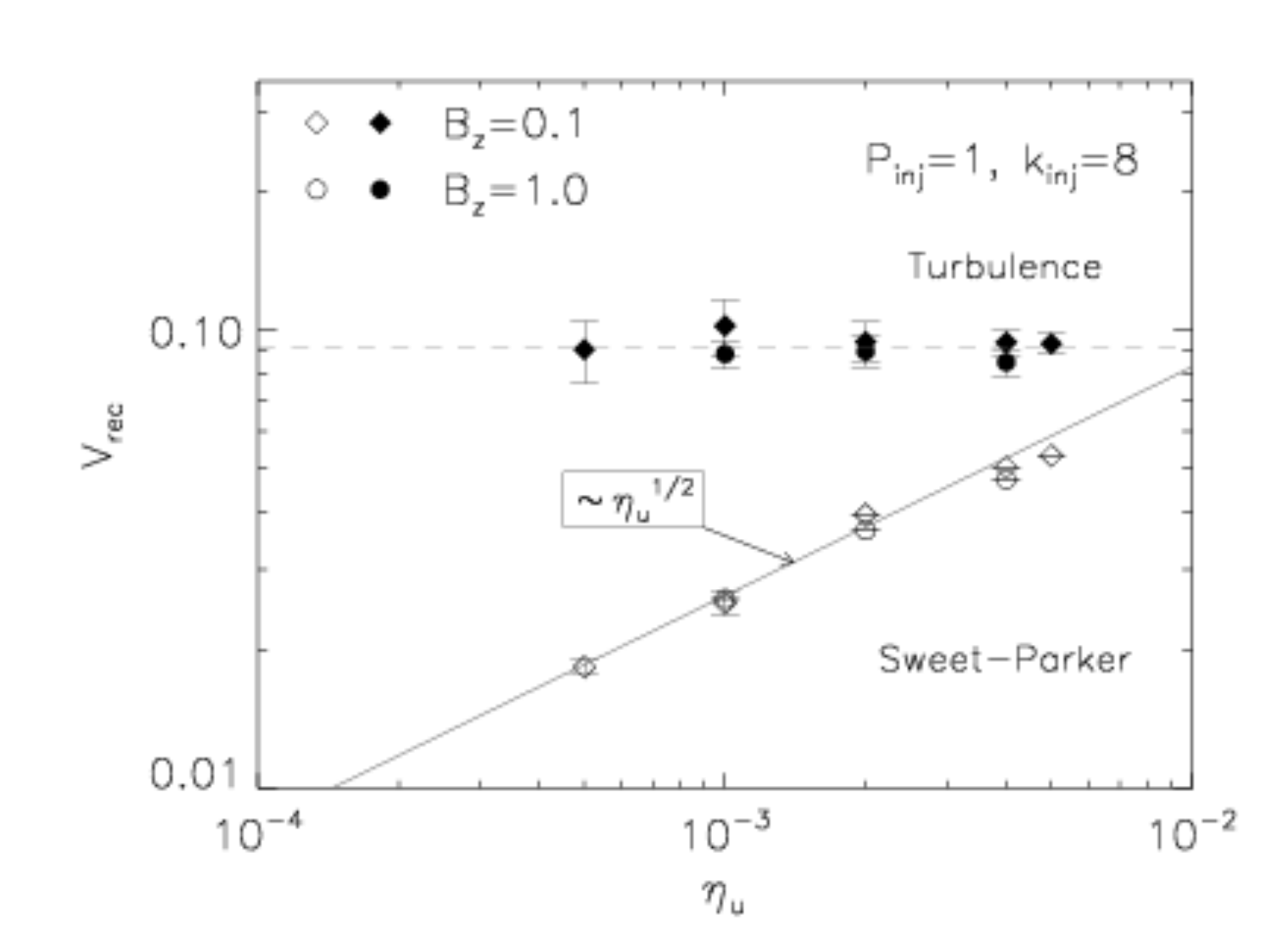}
\caption{Reconnection speed versus resistivity.  We show the reconnection speed
plotted against the uniform resistivity of the simulation for an injection
wavenumber of 8 and an injected power of one.  We include both the laminar
reconnection speeds, using the hollow symbols, fit to the expected dependence of
$\eta_u$, and the stochastic reconnection speeds, using the filled symbols.  As
before the symbol sizes indicate the uncertainty in the average reconnection
speeds and the error bars indicate the variance.  We included simulations with
large, $B_z=1$, and small, $B_z=0.1$, guide fields. \label{ueta_dep}}
\end{figure}

In Figure~\ref{ueta_dep} we plot the results for fixed input power and scale,
while varying the background resistivity.  In this case $\eta$  is taken to be
uniform, except near the edges of the computational grid where it falls to zero
over five grid points.  This was done to eliminate edge effects for large values
of the resistivity. We see from the Figure~\ref{ueta_dep} that the points for
laminar reconnection scale as $\sqrt{\eta}$, the expected scaling for
Sweet-Parker reconnection.  In contrast, the points for reconnection in a
turbulent medium do not depend on the resistivity at all. In summary, we have
tested the model of stochastic reconnection in a simple geometry meant to
approximate the circumstances of generic magnetic reconnection in the universe.
Our results are consistent with the mechanism described by LV99.  The
implication is that turbulent fluids in the universe including the interstellar
medium, the convection zones of stars, and accretion disks, have reconnection
speeds close to the local turbulent velocity, regardless of the local value of
resistivity.  Magnetic fields in turbulent fluids can change their topology on a
dynamical time scale.

In Kowal et al. (2009) we also studied the dependence of the reconnection on the
anomalous resistivity, which increases effective resistivity for high current
densities. The anomalous resistivity can be used as a proxy for plasma effects,
e.g. collisionless effects in reconnection. While it enhances the local speed of
individual reconnection events results in Kowal et al. (2009) testify that the
total reconnection rate does not change. In addition, the study of numerical effects
is presented in Lazarian et al. (2011). Numerical resistivity decreases with increase of the numerical resolution. Therefore, if the numerical effects were influencing
the reconnection rate, the increase of the resolution would decrease the reconnection rates. This is not what is seen in simulations (Lazarian et al. 2011).
If anything, the reconnection rates slightly increase with the increase of resolution
which is due to the fact that the turbulence proceeds to smaller scales. As a result, field wandering over smaller scales is available and therefore the outflow region
gets slightly thicker as the numerical viscosity decreases.

Finally, it is important to give a few words in relation to our turbulence
driving. We drive our turbulence solenoidally to minimize the effects of
compression, which does not play a role in LV99 model. The turbulence driven in
the volume around the reconnection layer corresponds to the case of
astrophysical turbulence, which is also volume-driven. On the contrary, the case
of the turbulence driven at the box boundaries would produce spatially
inhomogeneous imbalanced turbulence
for which we do not have analytical predictions (see discussion of such
turbulence in Beresnyak \& Lazarian 2009). We stress, that it is not the shear
size of our numerical simulations, but the correspondence of the observed
scalings to those predicted in LV99 that allows us to claim that we proved that
the 3D reconnection is fast in the presence of turbulence.

\section{Spontaneous onset of Turbulent Reconnection in Laminar System }

Macroscopically laminar systems can spontaneously transition toward unsteady regimes characterised by faster rates and presenting features that are closely related to the turbulent reconnection regimes described above.

The long recognized natural pathway for reconnection in laminar systems is the Sweet-Parker (SP) regime~\citep{sweet,parker}. In SP reconnection a thin and elongated current layer  forms separating two areas of opposing magnetic polarity. The plasma and the field  flows towards the current layer in the direction normal to the elongated side and outflows along the axis of the current.  Figure~\ref{flow1}-a shows the typical flow pattern.

The SP current layer, like any other current layer, is in itself unstable to the tearing mode \citep{biskamp-book, biskamp-viola} producing secondary islands. The presence of the flow pattern and the finite length of the layer prevent a direct application of the textbook analysis of the tearing mode. The first studies of the instability of SP layers were reported by \citet{bulanov} and are summarised in recent textbooks \citep{biskamp-book, biskamp-viola}. The conclusion was that for aspect ratios of the SP layer (length of the current channel, $\Delta$ over its width $\delta$) exceeding approximately 100, the SP layer becomes unstable. The aspect ratio of a SP layer is directly determined by the resistivity in the system. Using the parametrization of the resistivity with the Lundquist number $S$, the aspect ratio is simply $\Delta/\delta=\sqrt{S}$~\citep{priest-forbes}. It follows that given a sufficiently low resistivity, corresponding to a Lundquist number in excess of $S=10^4$, a SP  will become naturally unstable to formation of secondary islands. 

Low resistivity was not easily accessible to simulations until recently. At low resolutions, the numerical resistivity exceeds the intended low resistivity preventing the exploration of truly low resistive regimes. However, in recent years truly low resistivity regimes became accessible thanks to modern numerical methods and state of the art computational facilities. Great progress has been made in understanding this spontaneous transition from laminar SP reconnection to an unsteady regime characterised by a random onset of secondary islands. 

The configuration of a SP layer is intrinsically 2D and requires a 2D analysis~\citep{ni10}. Furthermore the SP layer is embedded in macroscopic configurations that further affect the stability and evolution~\citep{schindler99}. However, a reduced 1D analysis~\citep{loureiro07} can still take into account  the effects of the flow present in the SP layer.  The flow has a stabilising effect~\citep{bulanov-flow} that prevents the onset for aspect ratios below approximately 100, but at higher aspect ratios (corresponding to higher Lundquist numbers) the instability is present. The growth rate of the secondary instability is found to scale as $\gamma \approx S^{1/4}$ and the fastest growing mode corresponds to a number of islands that scales  as $S^{3/8}$~\citep{huang10}. The power is positive: the growth rate of the secondary instability is faster for lower resistivity. Additionally as the resistivity is lowered and the aspect ratio is increased, the number of magnetic islands increases.  The emerging picture from linear theory is then clear:  as the Lundquist number is increased the SP layer becomes progressively more elongated, with a higher aspect ratio. The more elongated the SP layer becomes, the larger the number of secondary islands becomes, and the faster they grow after forming. A clear indication of transition towards a turbulent regime like that covered in the previous sections. 

The prediction of the linear theory with respect to the number and growth rate of secondary islands have been tested and verified in carefully designed simulations~\citep{samtaney09, cassak09, bhattacharjee09} . 

Similarly, the threshold for onset (S approximately $10^4$) has been also confirmed~\citep{skender}. Additionally, the simulation studies have proven that turbulence, if present, has a direct impact on the process. The presence of pre-existing fluid turbulence changes the onset of the secondary island instability~\citep{skender,loureiro09}. A SP layer still forms also in presence of moderate levels of turbulence but it is more prone to become unstable.  The stronger seed reduces the threshold for the onset and the system transitions more quickly to an unsteady regime of reconnection characterized by multiple islands and reconnection sites. Indeed in all simulations the instability of the SP layer is seeded by numerical noise. Spectral methods avoid such noise completely and inedeed the instability cannot develop~\citep{ng11}, unless a seed in the form of a low level of turbulence is added explicitly.

An important consequence of the formation of secondary islands and the onset of unsteady reconnection is the acceleration of the reconnection process~\citep{lapenta08}.   Reconnection is usually defined as fast when two conditions are met. 

First, the reconnection rate should be independent of the mechanism that allows the decoupling of the plasma (and specifically the electrons in the case of advanced two fluid and kinetic models). In resistive MHD, the mechanisms allowing such decoupling are resistivity and, possibly, viscosity. With the onset of secondary islands, simulations observe not only an increased rate of reconnection but careful studies varying the resistivity in the system have shown that the rate becomes insensitive or even independent of resistivity altogether~\citep{lapenta08, loureiro09, huang10,uzdensky10}. 

Second, for reconnection to be fast in absolute terms, its rate must be a significant fraction of the local Alfv\'en speed measured with the density and magnetic field strength at the entry of the reconnection region $v_{A,in}$.   For a reconnecting Harris sheet, \citet{skender} report a rate of approximately $\Delta \dot{\Psi}/v_{A,in}B_{in}\approx .015$ with peaks of twice that value. Note that the value reported by \citet{skender} is computed with a different normalisation Alfv\'en speed based on the peak rather than the inflow density that is a factor of 10 smaller. Here we prefer to  translate the result in terms of to the actual Alfv\`en speed of the inflowing plasma for better comparison with the other litarature. For force free equlibria, the peak reconnection rate also nearly reaches $\Delta \dot{\Psi}/v_{A,in}B_{in}\approx .015$~\citep{skender}. Similarly, in the case of the reconnection induced by the coalescence of two magnetic islands, \citet{huang10} report a reconnection rate of $\Delta \dot{\Psi}/v_AB_0\approx .01$, in agreement with the case of the Harris sheet. The reconnection rate of unsteady MHD resistivity is therefore somewhat lower than that typical of the fastest kinetic regimes (peak rate of $\Delta \dot{\Psi}/v_{A,in}B_{in}\approx .1$~\citep{birnGEM}) but is still very strong.

Many of the properties of the fast reconnection process following the onset of the secondary islands instability can be understood with a simplified approach based on the properties of the tearing mode~\citep{bhattacharjee09}. For an initial current sheet  with a  magnetic field profile  
\begin{equation}
{\bf B}(z)=B_0 \tanh(z/a){\bf \hat{x}}
\label{harris}
\end{equation}
 the linear theory based on reduced MHD with uniform resistivity $\eta$ predicts a maximum growth rate~\citep[p.242]{schindler-book}:
\begin{equation}
\gamma_{max}a/v_A=0.623 \cdot (\mu_0a v_A/\eta)^{-1/2} 
\end{equation}
 corresonding to the wave number:
 \begin{equation}
 k_{max}a=1.358 \cdot (\mu_0a v_A/\eta)^{-1/4}
 \end{equation}

In the case of the stability of a SP layer, one can simply use the same result but consider that the initial state is itself the SP layer, with a thickness that is proportional to the square of the resistivity, $a=\eta^{1/2}$. It immediately follows that the fastest growing mode of the secondary island instability scales as $k \propto \eta^{-3/8}$ and the corresponding growth rate as $\gamma \propto \eta^{-1/4}$. These simple estimates agree remarkably with the more in depth analysis by \citet{loureiro07}. 

The argument can be carried further in the non-linear regime to estimate the non-linear reconnection rate during the subsequent growth of the islands. In the case of the tearing mode, the island growth is governed by the Rutherford equation~\citep{rutherford73}, stating that the size of the magnetic islands ($w_I$) grows in time linearly (rather than exponentially). The temporal derivative of the size of a magnetic island  is 
$ \dot{w_I}\propto \eta\Delta^\prime/\mu_0$~\cite[p. 76]{kadomtsev}, where, for the initial configuration  assumed above,
  \begin{equation}
\Delta^\prime=\frac{2}{a}\left(\frac{1}{ka}-ka\right)
 \end{equation}
At long wavelengths, where $ka<<1$, $\Delta^\prime\propto a^{-2}$, and assuming again the initial thickness to be that of the SP layer, the Rutherford equation  leads to a rate of island growth independent of resistivity,  $\dot{w_I}\propto \eta^{0} $.

Note that the dependence on $k$ is not considered in the scaling argument because the secondary islands are observed in the simulation to coalesce and the value of $k$ is expected  to settle on a value constrained more by the macroscopic system rather than by the fastest growing linear mode.

\section{Role of the flow pattern in unsteady reconnection}

A consequence of the onset of the secondary island instability is the profound modification of the plasma flow patterns in the reconnection region. To investigate the flow, a specific simulation is used. 
The initial state is a Harris sheet with magnetic field given by eq.~(\ref{harris}), uniform initial temperature corresponding to $v_s/v_A=1$ and density 
\begin{equation}
\rho=\rho_0 /\cosh^2(z/a)+\rho_b
\end{equation}
with $\rho_b= \rho_0/10$. The evolution is followed with visco-resistive MHD, using the FLIP3D-MHD code~\citep{jub-flipmhd}.

The simulations have  uniform resistivity corresponding to a Lundquist number, $S=\eta_0av_A/\eta=10^4$ and a viscosity corresponding to a Reynolds number, $R=\rho_0 v_A a/\eta=10^4$. All results are presented in normalised units where the magnetic field is normalised to $B_0$, the density to $\rho_0$, time with the Alfv\'en time $\tau_A=a/v_A$, space with the initial current sheet thickness $a$ and velocity with the Alfv\'en speed $v_A=B_0/(\mu_0\rho_0)^{1/2}$. The simulation box has sides $L_x/a=240$ and $L_z=60$. The code is 3D but the $y$ dependence is suppressed.  

\begin{figure*}[t]
\vspace*{2mm}
\begin{center}
a) $v_a t=84$\\
\includegraphics[width=.8\textwidth]{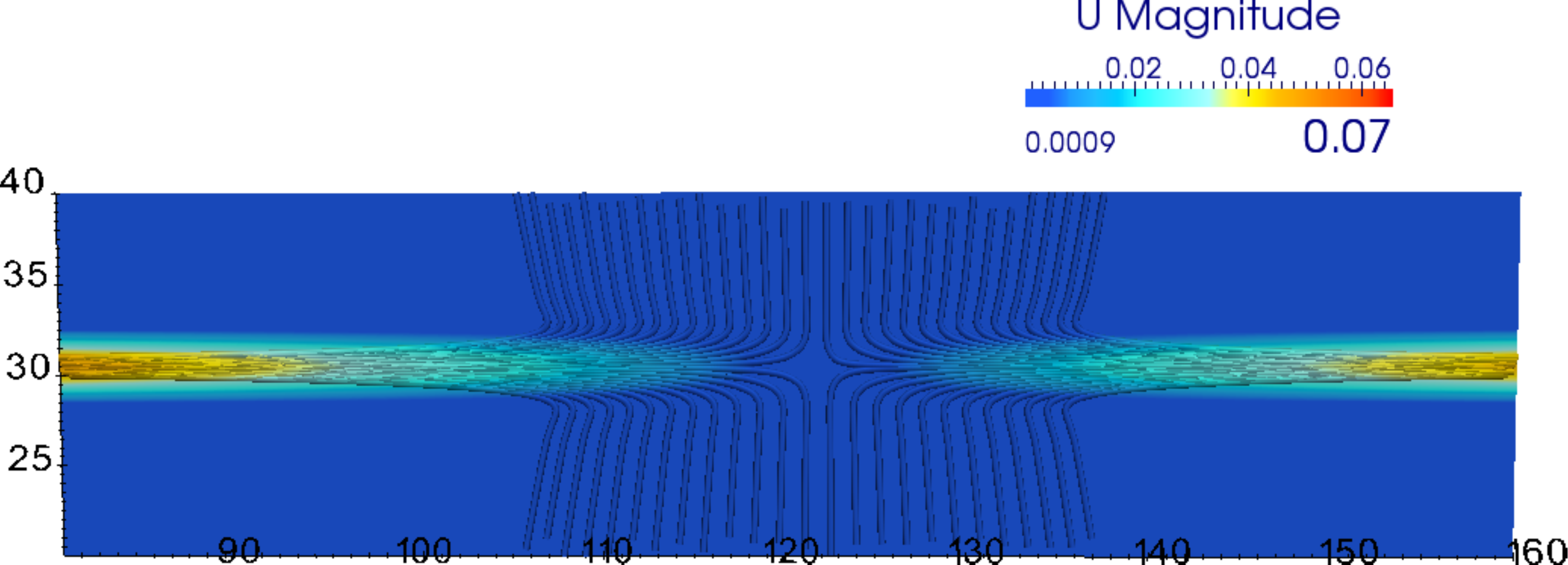}\\
b) $v_a t=208$\\
\includegraphics[width=.8\textwidth]{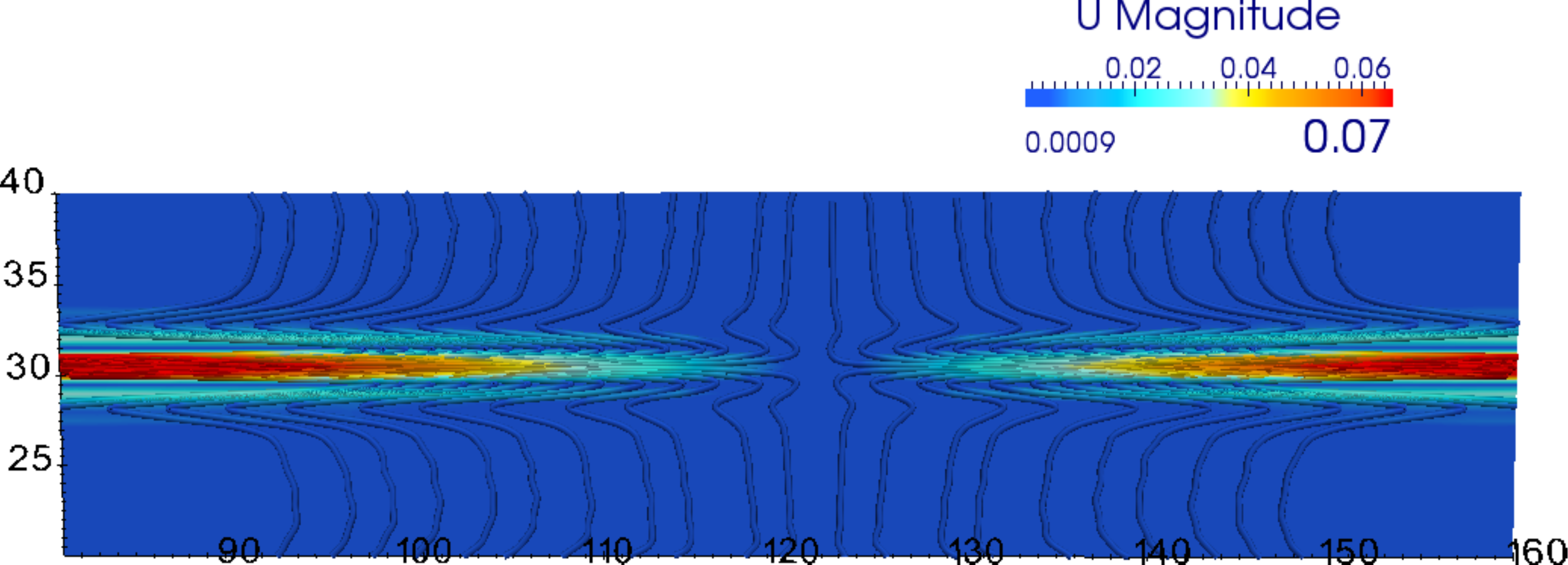}\\
c) $v_a t=308$\\
\includegraphics[width=.8\textwidth]{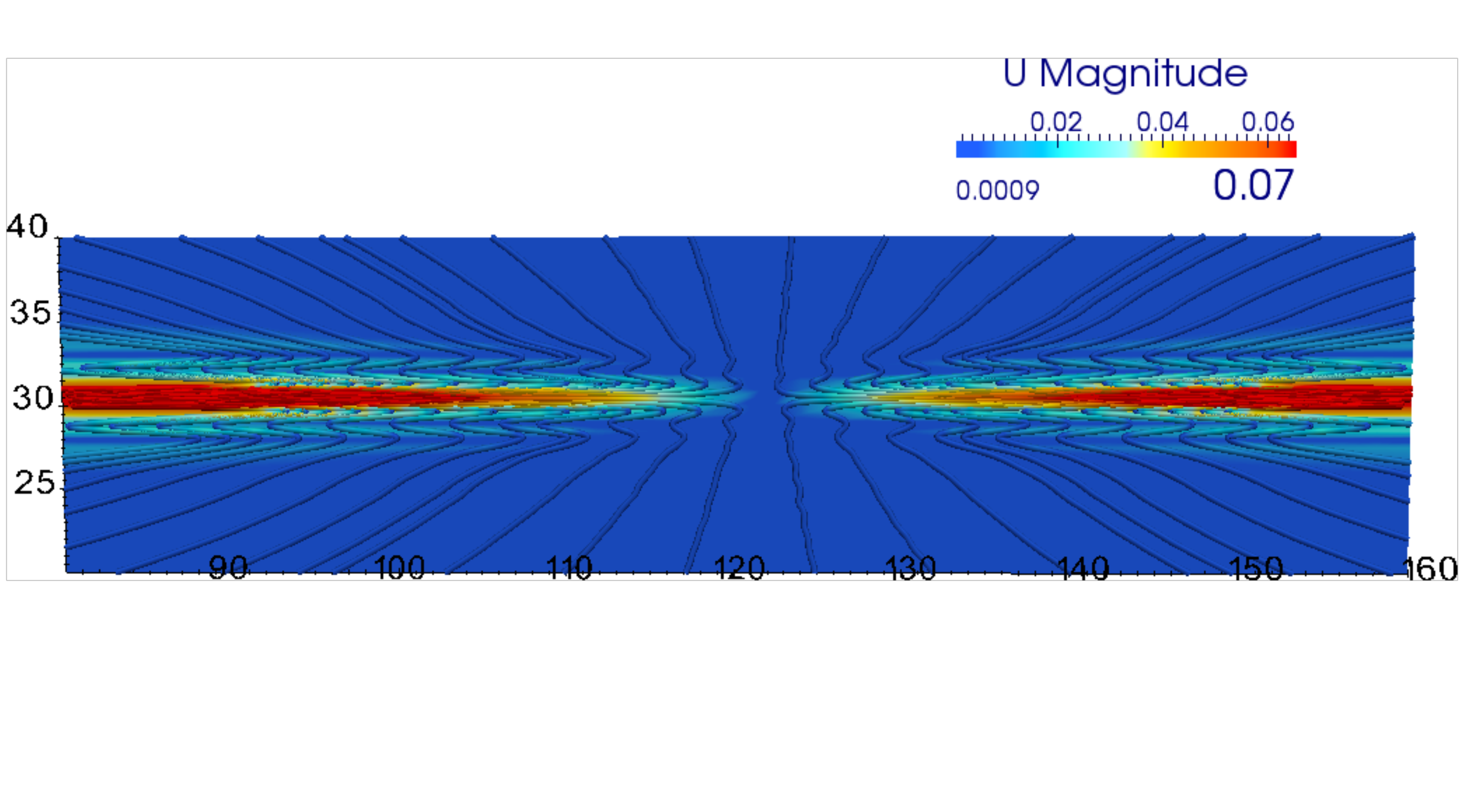}\\
\end{center}
\caption{Early evolution. The flow lines are shown superimposed over a false colour representation of the flow speed. The SP layer first forms and the flow pattern around it changes its nature. Three times are shown. Blow up of the central region, the full box is 240x60.}
\label{flow1}
\end{figure*}

During the SP regime, the flow is primarily directed normal to the sides of the SP layer. In the two inflow sides, the flow is vertical, and in the two outflow sides it becomes horizontal, veering within the SP layer. Figure~\ref{flow1}-a shows such a configuration for the present simulation. 
The actual flow in the simulation is just as the typical textbook SP cartoon~\citep{biskamp-viola}.

As the simulation evolves, the flow pattern changes and becomes primarily focused near the separatrices. This is a typical feature of the flow in presence of magnetic islands due to the tearing instability. Figure~\ref{flow1}-b and Fig.~\ref{flow1}-c demonstrate the transition. A x-point is forming in the center as a consequence of the secondary island instability. The flow pattern becomes now just like the textbook cartoon of the flow near an x-point in the tearing mode~\citep{biskamp-viola}.

\begin{figure*}[t]
\vspace*{2mm}
\begin{center}
a) $v_a t=332$\\
\includegraphics[width=.8\textwidth]{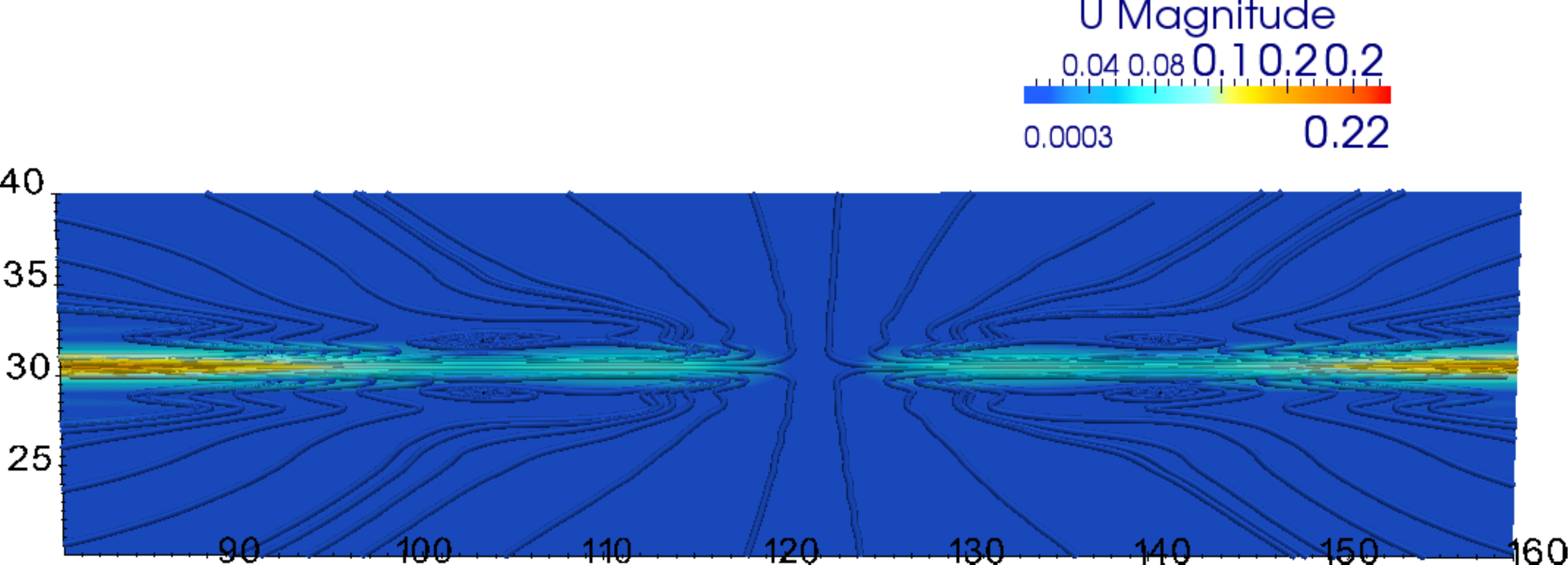}\\
b) $v_a t=352$\\
\includegraphics[width=.8\textwidth]{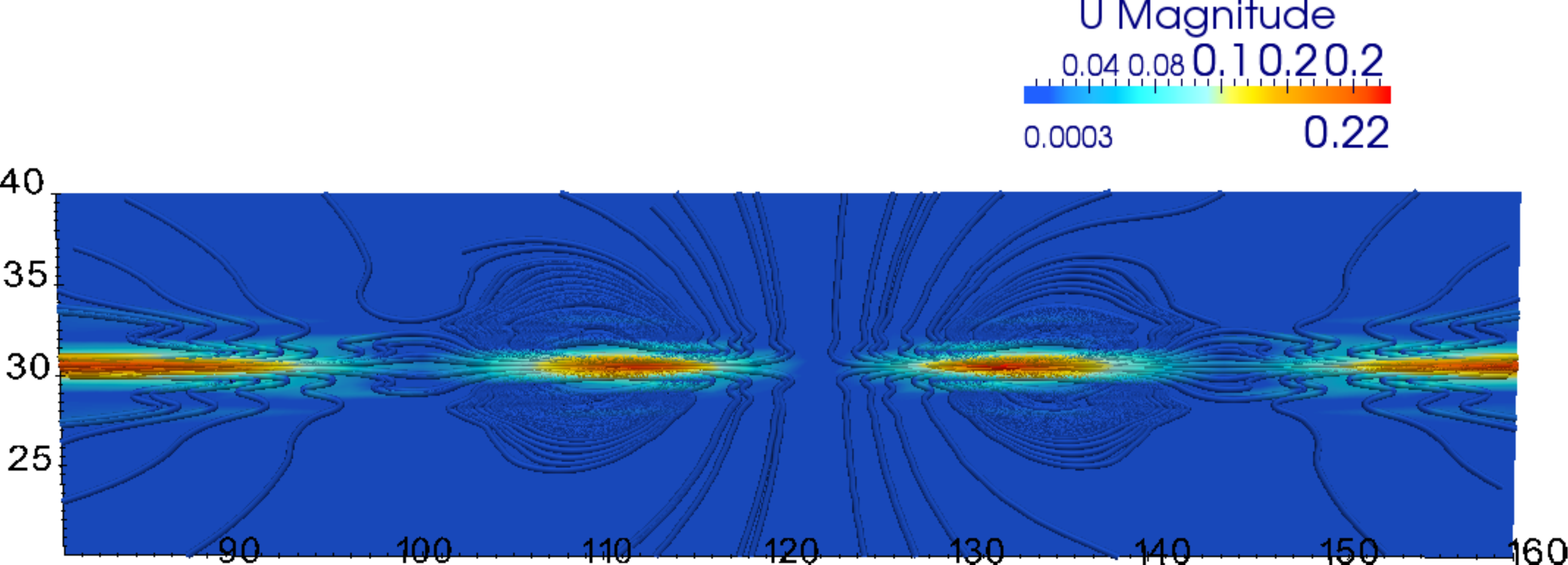}\\
c) $v_a t=368$\\
\includegraphics[width=.8\textwidth]{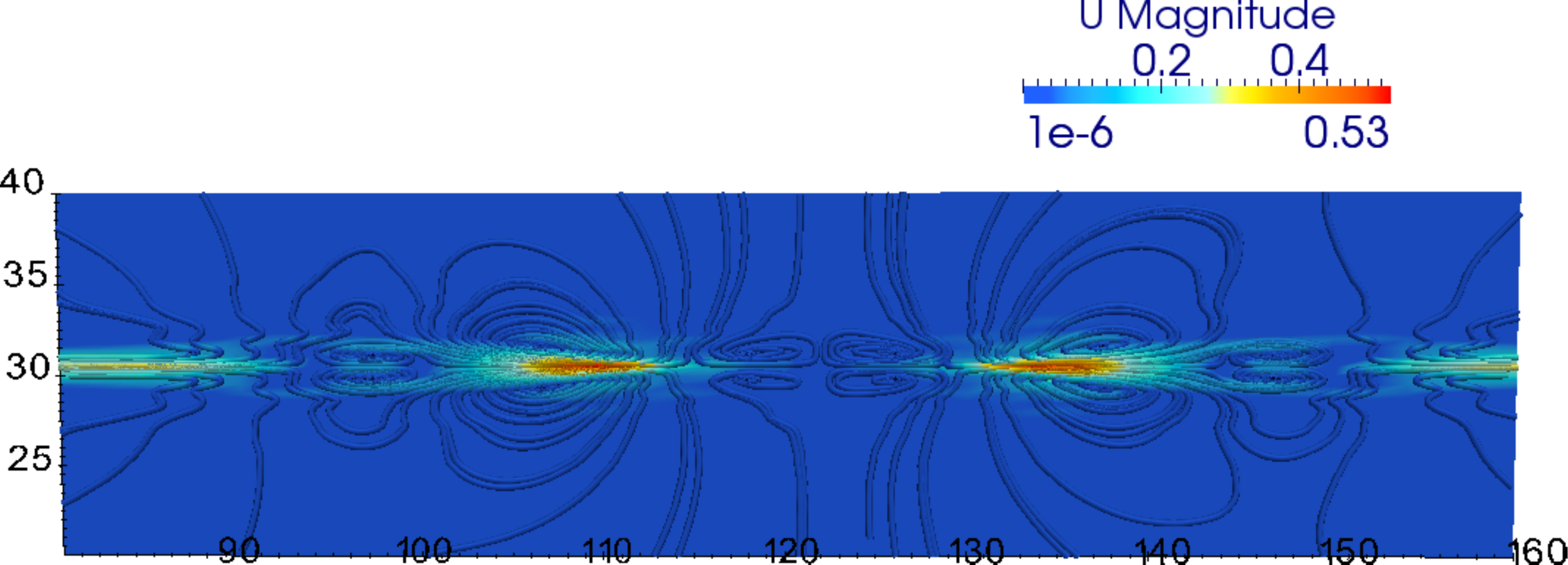}\\
\end{center}
\caption{Later evolution. The flow lines are shown superimposed over a false colour representation of the flow speed. The SP layer is destabilised by the secondary island instability. Three times are shown. The first two have the same color scale, but the third has higher values to avoid saturation. Blow up of the central region, the full box is 240x60. }
\label{flow-late}
\end{figure*}

Still later times are shown in Fig.~\ref{flow-late}. After the onset of the secondary island instability progressively more secondary islands and x-points are formed. The new formed islands coalesce with pre-existing islands and migrate with the flow towards the two ends of the simulation box~\citep{lapenta08}. The present simulation uses periodic boundary conditions but the same series of events develops also with open boundary conditions~\citep{skender, bettarini10} with the only difference that the flow carries the formed islands outside the open boundaries. 

The presence of multiple islands causes the flow to form characteristics loops. In a previous paper~\citep{lapenta08}, the name conveyor belts was used to explain the role of such flow loops.  The flow assumes a pattern where the flow lines that come out of the reconnection process bend and return towards the same reconnection region. The effect is similar to that of recursive reconnection~\citep{parnell} observed in magnetic skeleton configurations representative of solar coronal configurations. In recursive reconnection the outflow of one reconnection site feeds into another. In the present simulation the same recursive process is also observed. In Fig~\ref{flow-late}-c the outflow from the x-point located at $x/a=115$ feeds the reconnection process at the x-point located at  $x/a=90$ (and the similarly by symmetry on the other side). But more conspicuously, the same reconnection x-point at  $x/a=115$  feeds itself with most flow lines going out of it and coming back forming a conveyor belt. 

\begin{figure*}[t]
\vspace*{2mm}
\begin{center}
a)\\
\includegraphics[width=.8\textwidth]{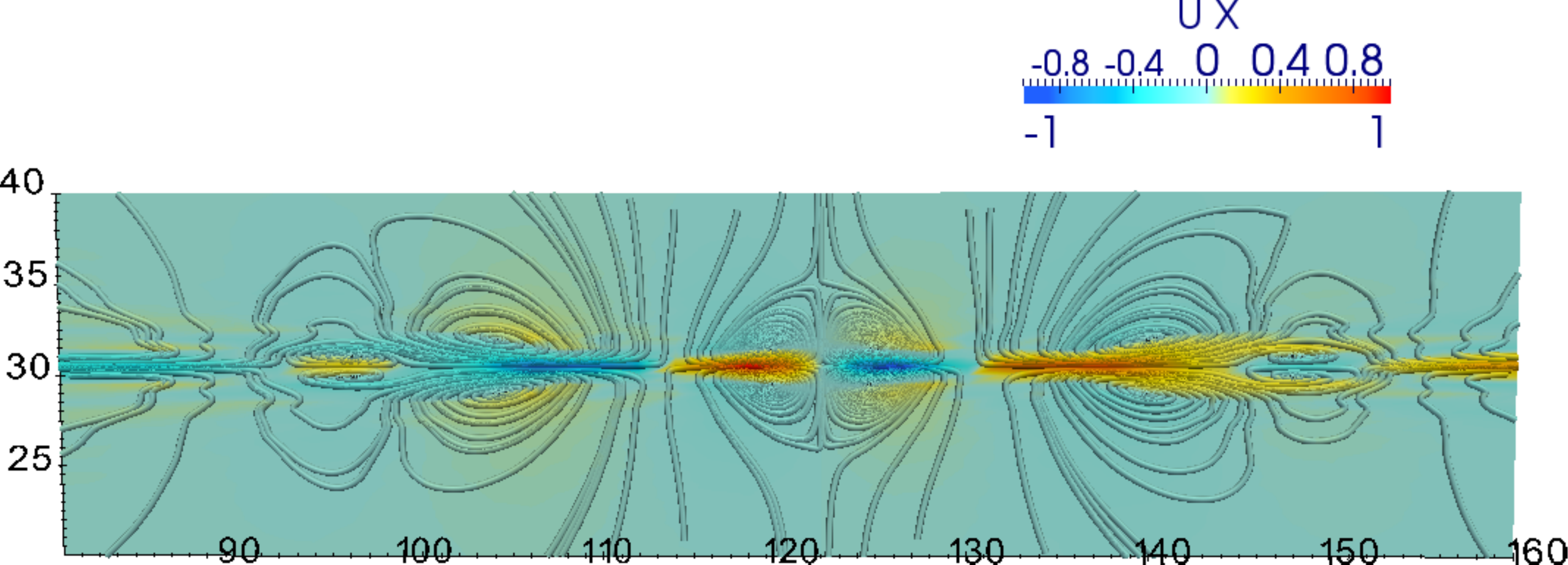}\\
b) \\
\includegraphics[width=.8\textwidth]{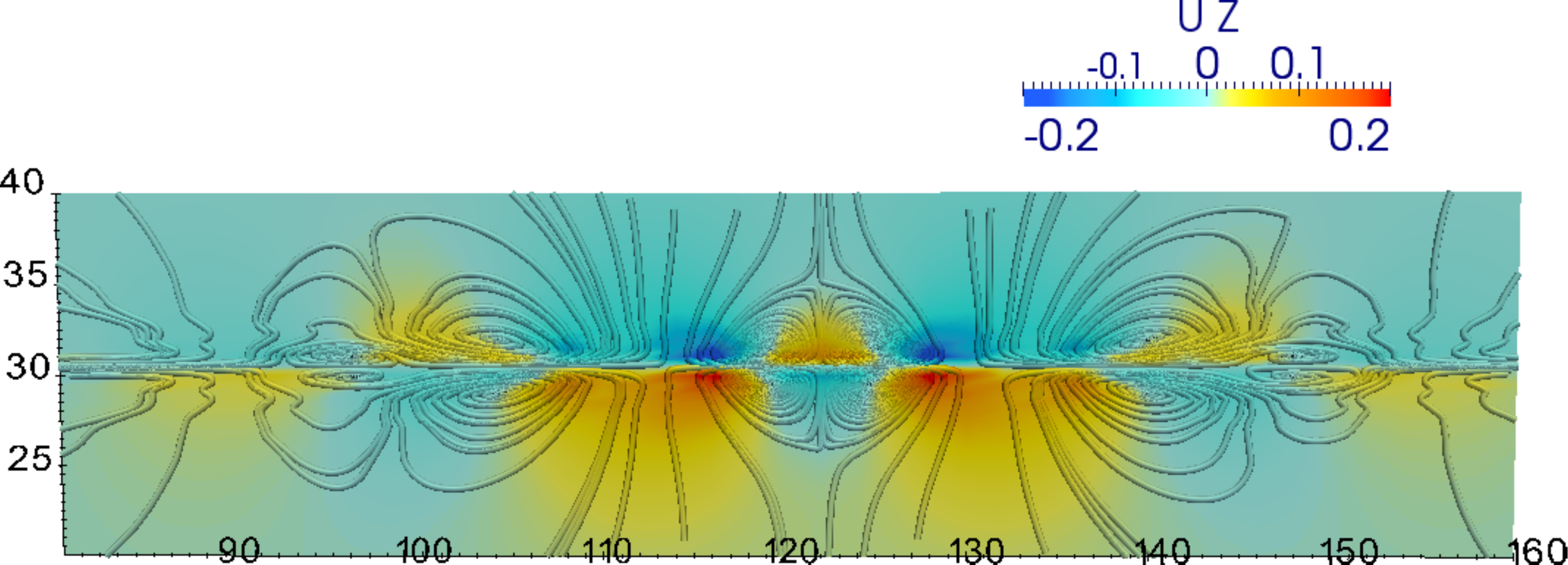}\\
\end{center}
\caption{Flow pattern at time $ t/\tau_A=375.5$. The outflow component $U_x$ is shown in panel a and the inflow component $z$ in panel b. Blow up of the central region, the full box is 240x60. }
\label{flow-value}
\end{figure*}

In accordance with the theoretical predictions above for reconnection to be fast the flow must become stronger in the unsteady regime. Comparing the scales in Fig.~\ref{flow1} with Fig.~\ref{flow-late}, the flow accelerates strongly. Figure~\ref{flow-value} shows separately the inflow component along $z$ and the outflow component along $x$. The outflow is at the Alfv\'en speed and the inflow is a significant fraction of it. At steady state the inflow speed equals the reconnection electric field and is by definition the reconnection rate. 

The concept of streamlines and in particular of conveyor belts should not be confused with actual recirculation of plasma. The flow pattern is distinct from the matter actually flowing. On those stepping into rivers staying the same other and other waters flow~\citep[DK22B12,Fragment 91] {heraclitus}. The flow can present the same pattern over a certain period but the plasma elements going through the flow are other and other. The streamlines are instant photos of the velocity field at a given instant. They are obtained here with the paraview software using the fourth-fifth order Runge-Kutta ODE solver that traces the lines everywhere tangent to the velocity field at that instant. But an actual fluid element feels at every instant the local speed at that instant. A moment later it feels another speed. Flow lines are not the same thing as particle trajectories. To analyse actual trajectories of fluid elements we use Lagrangian markers. Lagrangian markers are real particles of the fluid that move with the local instantaneous velocity just as a physical fluid would. These are not kinetic particles moving with their peculiar speed but rather fluid elements moving with the local flow speed. The FLIP3D-MHD code is based on a dual grid-particle representation that includes the presence of Lagrangian markers~\citep{jub-flipmhd}.

\begin{figure}[t]
\vspace*{2mm}
\begin{center}
\includegraphics[width=\columnwidth]{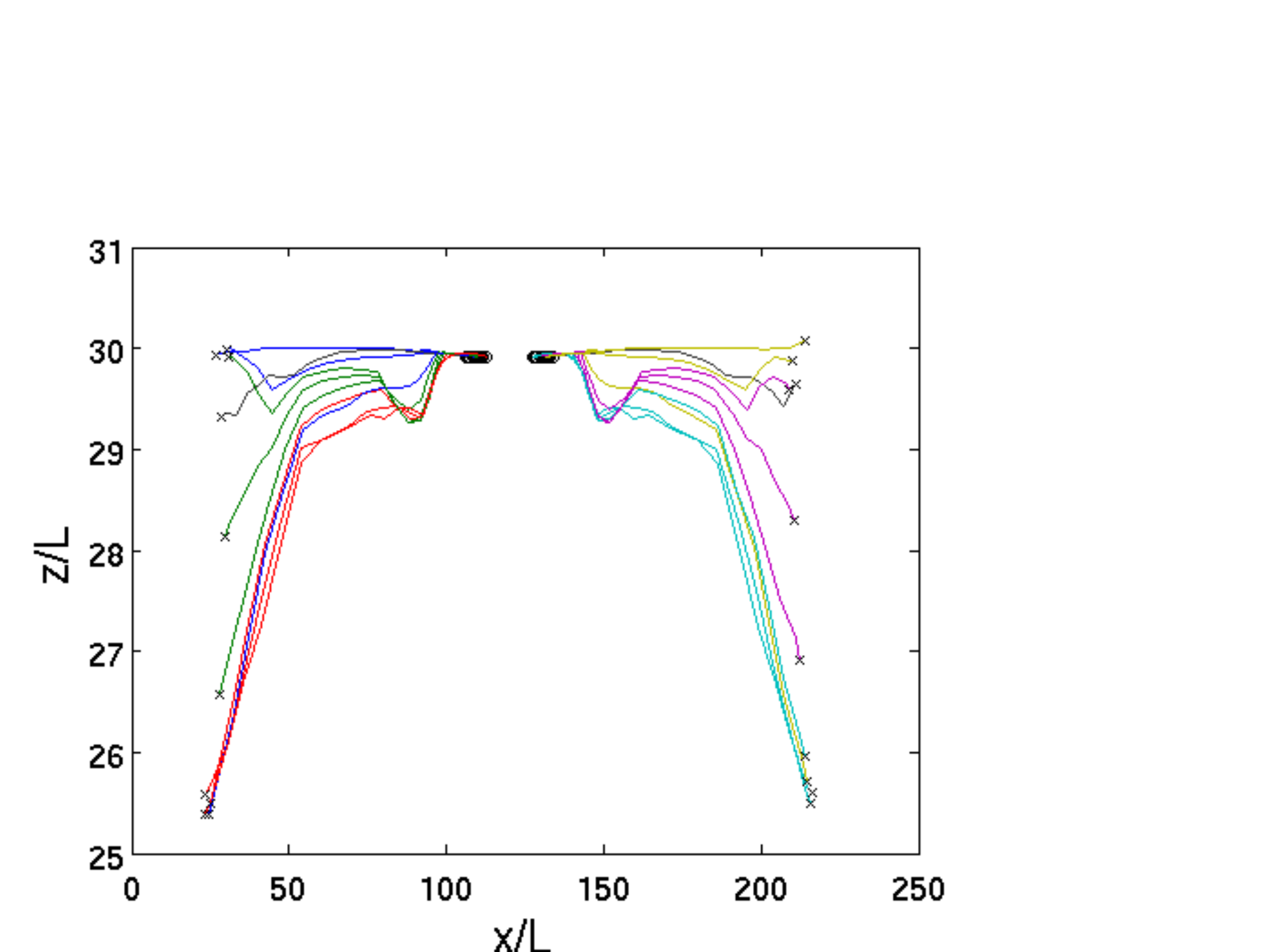}
\end{center}
\caption{Trajectory of Lagrangian markers. Miltiple trajectories are shown in different colours. Each trajectory starts at a circle and ends at a cross. The initial positions for the aprticles are equallys paced  between $x-L_x=\pm7L$ and $x-L_x/2=\pm14L$ on either side from the center of the x axis and at vertical position $z/L=29.9167$, just a little below the central neutral line. }
\label{tracce}
\end{figure}

Figure \ref{tracce} shows a selected number of trajectories of Lagrangian markers. The initial starting point at time $t=0$ in the simulation is chosen appropriately to identify particles that will go through multiple reconnection sites. Some trajectories only pass reconnection once: those are particles that are drawn toward the neutral line ($z/a=30$) and then ejected away from it (reddish trajectories on the left and light cyan on the right). 
However, there are other classes of trajectories (green on the left and magenta on the right) that are first drawn towards the neutral line and after being pushed away, come back a second time towards the neutral line and then are finally ejected away from it. These are particles that are captured in the conveyor belts and are pushed  towards a reconnection site and then away from it and then towards another one again. Note, however, that these trajectories are still monotonic in the horizontal direction: the same particle does not turn back to go towards the same x-point. For example on the left of center, the markers first take part to reconnection at the site located near $x/a=115$  and then at $x/a=90$. No particle is detected to  go through the same reconnection site twice. Just as the citation by Heraclitus suggested.

\section{Discussion}

A magnetic field embedded in a laminar perfectly conducting fluid preserves its topology for all time (Parker 79).  Although ionized astrophysical objects, like stars
and galactic disks, are almost perfectly conducting, they show indications of
changes in topology, ``magnetic reconnection'', on dynamical time scales (Parker
1970, Lovelace 1976, Priest \& Forbes 2002).  Reconnection can be observed
directly in the solar corona ( Innes et al 1997, Yokoyama \& Shibata 1995,
Masuda et al. 1994), but can also be inferred from the existence of large scale
dynamo activity inside stellar interiors (Parker 1993, Ossendrijver 2003).
Solar flares (Sturrock 1966) and $\gamma$-ray busts (Fox et al. 2005, Galama et
al. 1998) are usually associated with magnetic reconnection.  Previous work has
concentrated on showing how reconnection can be rapid in plasmas with very small
collisional rates (Shay et al. 1998, Drake 2001, Drake et al. 2006, Daughton et
al. 2006), which substantially constrains astrophysical applications of the
corresponding reconnection models\footnote{ We note that if magnetic reconnection is slow in some astrophysical
environments, this automatically means that the results of present day numerical
simulations in which the reconnection is inevitably fast due to numerical
diffusivity do not correctly represent magnetic field dynamics in these
environments. If, for instance, the reconnection were slow in collisional media
this would entail the conclusion that the entire crop of interstellar,
protostellar and stellar MHD calculations would be astrophysically irrelevant.}.

In comparison, our results point out to an attractive universal astrophysical solution for the magnetic reconnection: magnetic reconnection in presence of turbulence is always fast independently of the collisional or collisionless state of plasmas. The
formal criterion of the applicability of theLV99 model based on the MHD approximation to realistic plasmas is discussed in Eyink, Lazarian \& Vishniac (2011).   

Our paper provides a connection between the LV99 theory and numerical simulations demonstrating the spontaneous onset of fast reconnection. Indeed,
according the LV99 model the reconnection in the initially laminar state is slow. The
outflow, however, incites turbulence in the system. This turbulence induces faster
reconnection and a wider outflow gets more turbulent. This should induce a positive
feedback process which results in the flare of reconnection. We believe that this
is what is the actual theoretical explanation of the spontaneous onset of fast 
reconnection observed in papers by Lapenta et al. 

The spontaneous onset of reconnection is likely to be the cause of the solar
flares. The flares require the accumulation of magnetic flux of opposite polarity 
prior to the flare. The accumulation means low reconnection rate, which is
readily achived for the observed mean level of turbulence in solar atmosphere (Eyink et al. 2011). As the flux is accumulated the effects that we described in this paper
are expected to come into place and a abrupt release of energy through magnetic
reconnection is expected. A similar process can take place during gamma
ray busts (Lazarian et al. 2002, Zhang \& Yan 2011). A relatively low level of 
turbulence is a prerequisit for the flares. In the presence of high amplitude turbulence, reconnection is going to be fast from the very beginning preventing
the accumulation of magnetic flux at the pre-flare stage. 

Observationally, predictions of the LV99 model for flares are in rough agreement
with the estimates of the thickness of the outflow regions obtained in observations (Ciravella \& Raymond 2008). In addition, the initiation of the reconnection by
Alfvenic perturbations predicted in LV99 was confirmed by observations in 
Sych et al. (2009). Direct observations of turbulent reconnection in other environments, e.g. in interstellar medium are not feasible at the moment. However,
the indirect evidence of LV99-type reconnection comes from the observations of
magnetic fields in molecular clouds and accretion disks. In fact, on the basis
of LV99 model Lazarian (2005) predicted the process of magnetic field removal
termed "reconnection diffusion". Numerical studies of the reconnection diffusion
(see Santo de Lima et al. 2010, 2011) deliver results on expected magnetic flux diffusion that are consistent with observations.

The caveat here is that the LV99 model is the 3D model and it appeals to the 3D wondering of magnetic field lines. We believe that the fact that the acceleration was observed also in 2D systems testify that the modification of LV99 model is applicable to the systems of reduced dimentionality. It is discussed in Eyink et al. (2011) that
the existence of the "rough" turbulent velocity field should induce the Richarson diffusion and therefore the effective magnetic field wandering. The corresponding
theory have not been developed for the 2D turbulence. However, results on the
acceleration of reconnection in the presence of turbulence (Loreiro ..., Kulpa-Dubel et al. 2010) suggest that the results of LV99 may carry over to 2D systems. However,
as turbulence in 2D and 3D is different, we feel that it does not worth rederiving the 
LV99 results for 2D, but it is better to concentrate on matching the quantiative results for the actual real world 3D reconnection as numerical simulations are getting sufficiently powerful to study the effects in 3D. This is the theme of our intended
next paper. 

Our paper sends a warning signal to a naive interpretation of astrophysical reconnection within laminar reconnection models. We showed that pre-existing
turbulence as well as turbulence induced by the process of magnetic reconnection itself can dramatically change the reconnection rates. Interestingly enough, the
small - scale reconnection events may be determined by small scale physics, but
the total reconnection rates may still be determined by turbulent level in the system.

\conclusions[Summary]
The results of the present paper can be briefly summarised in the following way:\\

1. The LV99 theory of fast turbulent reconnection and more recent results on the spontaneous onset of turbulent reconnection are related to each other. We claim
that the flares of reconnection observed in simulations are due to the feedback
by turbulence induced by the outflow on the reconnection process. 

2. The numerical results suggest that flares of magnetic reconnection, e.g. 
explaining solar flares, may be explained on the basis of pure resistive MHD. 

3. Our work testifies in favor of a widely applicable model of astrophysical reconnection that is independent of subtle plasma effects. At the same time
it send a warning to attempts to model astrophysical reconnection without
including effects of turbulence.

\begin{acknowledgements}
The research of AL is supported by
 the NSF grant AST 0808118 and the Center for Magnetic Self Organization in
Laboratory and Astrophysical Plasmas (CMSO).
\end{acknowledgements}

\bibliographystyle{copernicus}



\end{document}